# Organizational and Socio-Technical Challenges in UAV Incidents: Evidence from a Practitioner Focus Group

*Full Paper*[1]

**George Grispos**
University of Nebraska-Omaha
ggrispos@unomaha.edu

**Joel Elson**
University of Nebraska-Omaha
jselson@unomaha.edu

**Austin Doctor**
University of Nebraska-Omaha
adoctor@unomaha.edu

**Suat Cubukcu**
Towson University
scubukcu@towson.edu

## Abstract

Unmanned Aerial Vehicles are now widely used across business, government, and recreational contexts, creating new challenges for incident response and digital forensics. While previous forensics research has largely focused on the extraction of technical data from UAV systems, minimal empirical work has examined how UAV incidents are handled in real-world settings or what challenges incident handlers face during this response. To address this gap, this paper reports findings from an in-person focus group with UAV and counter-UAV practitioners from industry and government organizations in the United States. Using qualitative analysis, several key challenges are identified, including situational awareness and airspace visibility, fragmented reporting and interorganizational coordination, forensic and attribution limitations, legal and policy gaps, and shortfalls in training and operational capacity. The research extends socio-technical incident response research to the UAV domain by providing practitioner-driven insight into UAV incident response and highlighting opportunities to strengthen incident response in this domain.

## Introduction

Unmanned Aerial Vehicles (UAVs) have transitioned from specialized military platforms to being used in commercial, government, and recreational settings. According to the Federal Aviation Administration (FAA), over 800,000 UAVs are registered in the United States (U.S.), with more than half licensed for commercial use (Federal Aviation Administration, 2025a). This adoption has enabled UAV integration into various domains including logistics, agriculture, emergency response, and infrastructure inspection (Shakhatreh et al., 2019). However, as UAV adoption has expanded, so too has the misuse cases, ranging from unauthorized surveillance to airspace incursions, increasing demand for effective UAV incident response (Mcmillan, 2025). Recent U.S. incidents illustrate the scale of this demand, including more than 2,800 unauthorized UAV incursions into restricted stadium airspace and over 350 incursions at military installations in 2024 alone (Daleo, 2025; United States House Oversight Committee, 2025).

UAV incident response refers to the structured procedures used by authorities to assess, manage, and investigate UAV-related incidents, including identifying UAVs and their operators, mitigating potential threats, and preserving physical and digital evidence (INTERPOL, 2020). Several organizations, including INTERPOL (2020) and the U.S. Cybersecurity and Infrastructure Security Agency (2025a) have published guidance on the handling, recovery, and investigation of UAVs involved in potential incidents. Complementing these best practices, academic research has focused on the foundations of UAV forensics, including flight log extraction, controller artifact analysis, telemetry parsing, and cloud data recovery (Salamh et al., 2021; Studiawan et al., 2023).

While previous research has documented UAV artifacts and developed valuable tools and approaches, they have largely been undertaken and evaluated in controlled settings. As a result, minimal empirical research examines how UAV incidents are managed in practice or what challenges incident responders face in real-world cases. In other domains, such as information systems, incident response requires coordination across technical tools, human decision-making, and organizational procedures, often involving stakeholders in different departments (Grispos et al., 2015; Line et al., 2014; Shedden et al., 2010; Werlinger et al., 2009). This interaction of technologies, people, and organizational structures suggests that incident response is inherently socio-technical in nature (Bostrom & Heinen, 1977; Dhillon & Backhouse, 2000).

Building on this idea, further investigation is needed to understand how UAV incident response is undertaken in practice and how incident handlers navigate organizational and investigative challenges during real-world incidents. To explore these issues further, a focus group was conducted with U.S. UAV and counter-UAV experts from industry and government organizations. Although the session was initially designed to examine the terrorist UAV threat, discussions organically expanded to include how UAV incidents are detected, managed, and investigated across organizational settings. These discussions were synthesized into a set of challenges that are reported in this paper.

This paper makes three contributions. First, it extends socio-technical incident response research to UAV incidents by examining the technical, organizational, and governance factors shaping UAV incident response. Second, it provides a practitioner-based insight into real-world UAV investigations, helping bridge the gap between technical forensics and operational practice. Third, it offers potential opportunities for improving reporting structures, forensic readiness and cross-organizational coordination in UAV incident response environments. The rest of the paper is structured as follows. The next section reviews related work and theory. This is followed by the methodology, the focus group findings, and a discussion of implications for research and practice. The paper then concludes and presents ideas for future work.

## Related Work

The growing use of UAVs has introduced new security and forensics challenges. Previous UAV security research has primarily examined vulnerabilities in communication channels, GPS spoofing, ground control systems, and the use of UAVs for malicious payload delivery (Shafique et al., 2021). Additional research has also examined counter-UAV technologies such as airspace monitoring, radio-frequency and radar sensing, and signal jamming (Lykou et al., 2020). When such defenses fail, a forensic investigation may be required to collect and analyze evidence from UAV hardware, controllers, cloud services, and telemetry data (Studiawan et al., 2023). Hence, existing UAV forensic research has concentrated on extracting flight logs, decoding storage formats, and recovering artifacts from platforms such as DJI drones (Kao et al., 2019; Mantas & Patsakis, 2019). Studies have also shown that mobile applications used to control UAVs can contain valuable forensic artifacts (Stanković et al., 2021). However, much of this forensic work has been conducted in controlled settings, rather than real-world investigations.

Evidence of real-world UAV misuse is mainly documented in homeland security reports, including cases involving prison contraband delivery, airport disruptions, and activity near critical infrastructure (Cybersecurity and Infrastructure Security Agency, 2025b; Department of Homeland Security, 2017). While these reports describe incident types, they provide limited insight into digital evidence handling, attribution, or end-to-end investigative workflows.

Within Information Systems, digital forensics research has focused on frameworks and organizational processes for collecting and preserving digital evidence (Reith et al., 2002). Similarly, related work on forensic readiness emphasizes the need for organizations to collect reliable evidence efficiently (Rowlingson, 2004). Hence, researchers have proposed forensic readiness frameworks for a variety of settings including corporate environments (Endicott-Popovsky et al., 2007) and cloud services (Trenwith

& Venter, 2013). However, these approaches largely focus on IT systems and do not fully address the challenges posed by UAVs.

Socio-technical systems theory argues that many security and operational failures arise from organizational coordination, communication, and decision-making issues, rather than purely technical weaknesses (Bostrom & Heinen, 1977; Dhillon, 2024). Incident response research similarly highlights the importance of aligning technical tools, human actors, and institutional structures. For example, Werlinger, et al. (2009) identify communication gaps, unclear roles, and decision bottlenecks as incident response barriers, while Grispos et al. (2015) emphasize stakeholder coordination, multidisciplinary expertise, and organizational learning as key capabilities. Security incidents also provide opportunities for organizational learning, by revealing weaknesses in both technical and socio-technical systems (Tøndel et al., 2014). Mechanisms such as post-incident reviews, root cause analysis, and cross-organizational knowledge sharing can all support this process (Grispos et al., 2015; Shedden et al., 2010). While previous research has focused on the technical aspects of UAV forensics, with most empirical work carried out in controlled or tool-focused settings, minimal work has examined how UAV incidents are investigated in practice or what challenges practitioners face. Moreover, there is limited research based on direct input from practitioners, such as focus groups, to understand real-world investigative experiences.

## Research Methodology

This study utilizes a focus group, a qualitative research method used to gather participant insights, refine concepts, and generate new research within a social context (Breen, 2006). This method is appropriate because it supports the exploration of tacit practitioner knowledge in emergent and underexplored domains (Ryan et al., 2014), particularly where operational knowledge is distributed across agencies and sectors. The focus group included fourteen subject-matter experts from U.S. industrial and governmental organizations. Government participants represented different parts of the federal law enforcement, intelligence, and homeland security communities, as well as state and local law enforcement agencies. Overall, the participants represented a diverse range of UAV and counter-UAV roles.

Invitations were distributed through professional networks and previous project collaborations. All participants provided informed consent prior to participation, and no personally identifiable information was collected during the discussion. The study received Institutional Review Board Exempt approval, and all discussions were conducted at the unclassified level in accordance with research ethics requirements.

The focus group discussion centered on several initial problems, including: the evolution of the terrorist UAV threat in the U.S., the risk factors associated with terrorist UAV deployment, target selection and critical infrastructure vulnerabilities, cybersecurity and risk factors, and barriers hindering effective counter-UAV efforts (Doctor et al., 2025). Although the initial focus group centered on terrorist threat scenarios, forensic and investigative challenges emerged organically throughout the session. Participants frequently described evidence-handling challenges, data acquisition barriers, attribution difficulties, and coordination issues affecting UAV incident response. As an exploratory study, these emergent themes were captured and analyzed as part of the inductive approach (Thomas, 2006). These insights form the empirical foundation for the findings presented in this paper.

The discussion was captured using handwritten notes by two members of the research team, while participants were also encouraged to take their own notes, which were collected at the end of the session. No audio or video recordings were made due to the exploratory and potentially sensitive nature of the discussion, as this was considered likely to limit open participation. Following the session, the researchers compared and combined their notes to improve completeness and reduce individual bias. All collected materials were anonymized prior to analysis.

The researcher and participant notes were then combined and analyzed inductively (Thomas, 2006), allowing key themes to emerge from frequent, dominant, or significant points in the data. Initial coding was carried out independently by two researchers, who then compared and discussed their codes to agree on a shared set of themes. This process supported the identification and consolidation of recurring issues across the dataset. While the absence of verbatim transcripts is a limitation, the use of multiple note sources and collaborative analysis helped improve the transparency and consistency of the findings.

This research has several limitations. First, the findings reflect the perspectives of a specific group of U.S. practitioners, who reported experiences rather than direct observations. Second, the focus group was constrained by security restrictions, which prevented recording the discussion. Third, given the nature of UAV incident response, the findings from this focus group with U.S. participants, may not generalize across national contexts and different legal and regulatory settings. Instead, the findings should be interpreted as an initial perspective into the socio-technical and organizational dynamics shaping UAV incident response, which could change as capabilities mature and these dynamics change over time.

## Findings and Analysis

The focus group identified five interconnected socio-technical and organizational challenges related to UAV incident response, *limited situational awareness and airspace visibility*, *fragmented incident reporting and data sharing*, *forensic constraints*, *legal and policy gaps*, and *capacity and training limitations*. Table 1 provides an overview of these key issues, including example participant observations and their potential impact on UAV incident response. The following sections discuss each challenge in more detail.

It must be noted that during the focus group, the participants used the terms "UAV", "drone", and "UAS" interchangeably. Although technically distinct (International Civil Aviation Organization, 2011), they are treated as equivalent in this study to reflect participant language.

| Challenge Area | Example Participant Quote(s) | Impact on UAV Incident Response |
|---|---|---|
| *Situational awareness and airspace visibility limitations* | "Do not know (cannot track) all UAVs in US airspace"; "hard to recognize drones from long distances"; inability to link remote IDs | Makes it harder to detect threats early, respond in real time, and maintain airspace awareness |
| *Fragmented reporting and data sharing* | "No centralized national database"; "no single agency has been designated as the point of contact for UAS threats" | Harder to identify patterns, coordinate across agencies, share intelligence, and learn from incidents |
| *Forensic limitations and attribution challenges* | "Operators often discard drones"; "cheap drones don't contain flight information"; "…assembly of drones completely off the grid" | Makes reconstruction, attribution, and reliable evidence collection more difficult |
| *Legal, policy, and enforcement gaps* | "Punishment is difficult to enforce"; limited authority outside TFRs; liability concerns affect counter-drone deployment | Slows response decisions, limits mitigation options, and complicates investigations |
| *Capacity, training, and resource constraints* | "Local law enforcement lacks training, resources, and authority"; response time limited; agencies struggle to keep pace with innovation | Leads to slower responses, weaker evidence preservation, and more reactive incident handling |

**Table 1. Summary of Identified UAV Incident Response Challenges**

### *Lack of Situational Awareness and Airspace Visibility*

Participants reported that agencies lack a complete picture of UAV activity in national airspace. Detection gaps were frequently mentioned, including that agencies "*do not know (cannot track) all UAVs in the US airspace*", and that it is "*hard to recognize drones from long distances*". These two challenges go beyond just the detection of malicious UAVs, adding that even when UAVs are visible, responders indicated that they "*cannot easily tell what a drone operator's intent is just by observing*". These statements align with FAA reports (2020) which indicate that the proliferation of small UAVs require new low-altitude surveillance and traffic management capabilities that go beyond traditional airspace monitoring systems. From an incident response perspective, this suggests that agencies may face limitations in maintaining continuous situational awareness of UAV activity.

Comments from the focus group suggested that certain UAV characteristics can further complicate detection and monitoring. For example, certain UAVs may emit weak or no detectable radio frequency or telemetry signals, with participants noting that "*drone forensic recovery is limited when drones are not emitting signatures*", indicating that limited signal emission can constrain both detection and subsequent investigative opportunities. This UAV characteristic also appears to complicate counter-UAV efforts, as some UAVs "*emit no signals that can be disrupted - those with mitigation tools can do nothing to stop it*". Together, these characteristics could reduce incident handlers' ability to implement timely intervention measures and recover usable evidence during incidents.

Participants also highlighted limitations in current identification infrastructure. One respondent noted "*there is no way to use ID numbers from remote ID on drones to link it in real time with detection systems*", suggesting that even where identification mechanisms exist, the lack of interoperability between detection platforms and identification systems may reduce real-time situation awareness. Interestingly the participants also suggested that increasing autonomous flight capabilities may further complicate situational awareness by reducing opportunities to observe operator behavior or identify control patterns, although specific mechanisms were not detailed.

## Fragmented Reporting, Data Sharing, and Institutional Coordination

The participants described the UAV incident reporting ecosystem as fragmented, with inconsistent procedures, data silos, and unclear responsibilities. They noted that "*there is no centralized national database to track UAS-related incidents*" and "*no centralized database of drone sightings and incidents*". In line with this thought, one participant added that "*only isolated events are reported; no large scale numerical analysis exists*". The absence of centralized incident repositories could have implications for organizational learning and coordinated response. Previous research shows that structured knowledge capture systems, such as incident response databases, help to identify patterns, improve response procedures, and help enhance cross-agency situational awareness (West-Brown et al., 1998). Without a centralized database, agencies may find it harder to spot patterns or predict threats.

This fragmentation also appears to extend into broader data sharing practices. As one participant noted, "*comprehensive datasets on UAS suspicious activity reports are not shared or synthesized*", which suggests institutional barriers exist that could limit cross-agency information sharing. Previous research similarly finds that organizations often struggle to document, disseminate and operationalize knowledge from security incidents, largely because of limited dissemination practices (Grispos et al., 2015).

Reporting pathways were also unclear. Participants questioned "*what is the right way to report and who do they report it to and what do they report?*", emphasizing that "*there is no one way to report a UAS*" and "*no single agency has been designated as the point of contact for UAS threats*". This lack of procedural clarity may further complicate the timely reporting of UAV incidents.

Reflecting this concern, one participant did suggest that "*standardized reporting forms should be created*". Previous research has shown that standardized reporting structures do indeed support effective incident response and organizational learning (Tøndel et al., 2014; West-Brown et al., 1998), suggesting that the current fragmentation may complicate coordinated investigations. In addition, standardized incident reporting forms can strengthen investigative efforts and data collection on information that provides value to the UAV incident response team in question (Grispos, 2016; Grispos et al., 2019).

## Forensic Limitations and Attribution Challenges

The participants highlighted significant challenges in relation to recovering usable forensic evidence and attributing UAV activity. In some incidents, it appears that physical evidence is unavailable because "*operators often discard drones, making it hard to trace them*". Such behavior suggests deliberate operator disposal practices, similar to counter-forensic strategies observed in other criminal contexts, such as the disposal of mobile phones to evade law enforcement detection (Keoghan, 2024).

Even when UAVs are recovered, certain evidence may be limited, as "*no fingerprints [are left] due to frequent handling once recovered*". Despite international guidance (INTERPOL, 2020), these statements suggest a potential gap between recommended evidence handling practices and those actually implemented in practice, possibly reflecting uneven training or awareness among incident handlers.

Technological developments were also identified as complicating investigations. For example, some low-cost UAVs lack onboard storage or telemetry, with participants noting that "*lots of cheap $50 drones don't contain any flight recording information*". UAVs without built-in logging make event tracing more difficult and limit an investigator's ability to undertake event reconstruction using stored flight data.

Open-source designs and do-it-yourself UAV construction were viewed as additional barriers, since "*open-source manufacturing allows for threat actors to assemble drones completely off the grid*", while "*3D printing increases the threat… no purchase history to track dangerous modifications*". The introduction of these distributed manufacturing ecosystems can make purchases harder to trace and weakens investigations that could rely on serial numbers, purchase records, or manufacturer data, which could force investigators to rely more on behavioral evidence rather than hardware tracking. These findings align with previous research on the UAV ecosystem, which shows that open-source designs, community-developed firmware, and accessible additive manufacturing lower technical barriers to entry for malicious actors while also reducing the ability to trace device origins and modifications (Grispos et al., 2025).

### *Legal, Policy, and Enforcement Gaps*

Participants suggested that legal and regulatory limitations are reducing the effectiveness of incident response efforts. For example, even when UAV pilots are identified, "*punishment is difficult to enforce*", with "*minimal fines and enforcement authority*". Limited legal authority may also restrict an incident handler's ability to intervene, seize UAVs and other equipment, or collect evidence. This reflects a broader challenge in technical governance, where laws often struggle to keep pace with rapid technological change (Moses, 2013). In UAV contexts, this legal lag may introduce uncertainty about when intervention is legally justified, and which agencies have enforcement authority.

A separate legal challenge identified is related to Temporary Flight Restrictions (TFRs), which are FAA airspace rules that temporarily limit flights over certain areas (Federal Aviation Administration, 2025b). Participants suggested that any legal authority to counter UAV activity remains limited outside of these restrictions, noting that "*there are no jurisdictions [or authorities] in place except to interdict [UAVs] when a TFR is in place*". However, the reliance on TFRs suggests a largely reactive governance model that could delay response actions and intervention efforts.

Liability concerns were also raised, with one participant noting that "*liability protection is going to be a part of the decision for dealing with counter-drone systems*". This suggests that incident handlers balance operational needs with legal risk when determining when and how to respond to an incident. However, such hesitations could reduce opportunities for timely intervention and evidence preservation.

### *Capacity, Training, and Resource Constraints*

Finally, participants emphasized training, resource, and preparation limitations. They noted that "*local law enforcement lacks training, resources, and authority to handle drone incidents effectively*". This concern was echoed further in the discussion, with another participant summarizing the issue as a "*lack of training, resources, and manpower to handle drone incidents effectively"*. West-Brown et al (1998) emphasize that preparedness, training, and established procedures are critical for effective incident response. However, as the focus group findings suggest, these limitations may reduce incident handlers' ability to recognize UAV threats and deploy appropriate measures.

Participants also emphasized the operational importance of rapid intervention, with one noting that "*response time is the most important factor in preventing*". During a follow-up, they indicated that rapid response is often not possible due to limited preparation and equipment.

Respondents also suggested that rapid UAV innovation is placing additional strain on agency capacity and training requirements. One participant observed that that "*the ability to use AI and low-cost models to learn how to use/modify a UAS makes it increasingly difficult for agencies to keep pace*", indicating that evolving technologies may outpace existing skills, equipment, and preparation levels. Others characterized the broader response systems as reactive, noting that "*…we only care when a major attack happens*". These observations suggest that resource limitations, training gaps, and preparedness shortfalls may hinder timely response, evidence preservation, and ultimately, organizational learning. Collectively, these findings show that UAV incident response challenges extend beyond individual technical limitations and reflect broader constraints in visibility, coordination, investigation, governance, and preparedness.

## Implications for Research and Practice

The focus group findings show that UAV incident response extends beyond technical artifact extraction and instead involves challenges spanning multiple socio-technical layers. Table 2 illustrates how these challenges emerge across *technical*, *organizational*, *institutional*, and *human* dimensions (Dhillon & Backhouse, 2000), indicating that response effectiveness depends on the interaction of technologies, operational processes, governance arrangements, and practitioner decision-making.

This extends socio-technical understandings of incident response (Dhillon & Backhouse, 2000; Werlinger et al., 2009) into the UAV domain, where previous research has largely focused on forensic tools and artifact recovery. These findings also highlight the need for research examining how forensic capabilities are integrated into operational workflows and used in real-world investigative settings (Line et al., 2014).

For practitioners, these findings highlight the need to move beyond tool-focused approaches and better integrate UAV evidence into investigative workflows, including coordination across teams. For the digital forensics community, the results highlight opportunities to improve how UAV data interpretation and integration with other evidence. For policymakers, the findings suggest a need for clearer standards around UAV incident response, particularly in relation to evidence handling and cross-agency coordination.

| Challenge | Technical Layer | Organizational Layer | Institutional Layer | Human Layer |
|---|---|---|---|---|
| *Situational awareness limitations* | Incomplete detection systems and limited interoperability | Limited shared monitoring processes | Unclear authority for integrated airspace oversight | Responders make decisions with limited visibility |
| *Reporting and data fragmentation* | No centralized incident databases | Inconsistent reporting and coordination practices | No clear reporting authority | Personnel unsure what and where to report |
| *Forensic and attribution constraints* | Limited onboard data; off-grid manufacturing | Uneven evidence collection practices | Weak traceability mechanisms | Greater reliance on inference over technical artifacts |
| *Legal and policy gaps* | Counter-UAV tools limited by authorization rules | Response shaped by jurisdictional boundaries | Limited enforcement authority; reactive measures (e.g., TFRs) | Operational hesitation due to legal uncertainty |
| *Capacity and training limitations* | Limited specialized equipment | Insufficient preparation and exercises | Uneven institutional investment in UAV readiness | Skills and readiness lag technological change |

**Table 2. Socio-Technical Mapping of UAV Incident Response Challenges**

From a practice perspective, the findings suggest that UAV incident response often lacks structured knowledge capture mechanisms that support organizational learning. In traditional incident response, centralized incident reporting systems and shared databases help improve coordination, pattern detection, and learning (West-Brown et al., 1998). Without comparable structures for UAV incidents, agencies may find it difficult to notice recurring threats, share actionable intelligence, and build forensic readiness (Fransen et al., 2015; Rowlingson, 2004).

The forensic and attribution challenges further indicate that current readiness strategies may not match the evolving UAV ecosystem. As UAV technologies move toward autonomous operations, low-cost platforms, and open-source construction, traditional post-incident evidence collection may become less reliable. Previous research (Rowlingson, 2004) shows that investigative practices must evolve with emerging technologies, highlighting the need for proactive readiness models that combine technical innovation, updated procedures, and specialized training.

Legal and regulatory frameworks were also found to shape UAV incident response. Previous research shows that governance structures influence decision-making authority, coordination, and response timing (Goodall et al., 2004; Werlinger et al., 2010). Similar constraints appear in UAV contexts, where regulatory limits may affect mitigation and investigative actions. For example, the reliance on TFRs suggests that UAV incident response may currently operate in a largely reactive governance model (Gillon et al., 2011), which can delay intervention and reduce opportunities for timely investigations.

Finally, the capacity and training gaps identified in the focus group indicate that workforce preparation remains critical for UAV incident response. Researchers have shown that response effectiveness depends on training, communication, and shared operational practices (Line et al., 2014; Werlinger et al., 2009). These findings suggest a need for stronger cross-organizational training, simulation exercises, and knowledge sharing mechanisms to further support coordinated incident management.

## Conclusions and Future Work

Drawing on focus group data, this paper examines UAV incident response through a socio-technical lens and shows that real-world UAV incident response activities extend beyond the technical forensic analysis reported in the literature. The findings indicate that issues with situational awareness, fragmented reporting structures, attribution challenges, legal ambiguities, and training gaps all influence how UAV incidents are detected, managed, and investigated. Overall, the results show that UAV incident response is a socio-technical system in which technologies, organizational processes, institutional constraints, and inter-agency coordination must work together to support effective incident management. Hence, the study contributes to an emerging area that has received minimal attention in the IS community.

From a practical perspective, the findings show that UAV incident response can be improved in several ways. Standardized reporting pathways may help make information more consistent and easier to track, while stronger knowledge-sharing practices could support organizational learning. Likewise, enhanced forensic readiness practices, including aligned tools, policies, and training, may also improve investigations outcomes. More broadly, stronger inter-agency coordination and clearer legal authority could help move from a reactive response to more proactive UAV incident management posture. As UAV technologies continue to emerge, addressing these socio-technical challenges will be important for supporting timely, coordinated, and evidence-supported UAV response practices.

There are several directions for future research. This research used a single focus group and U.S.-centric perspective, which may not capture broader organizational or international practices. Hence, future work should examine UAV incident response across more stakeholders and an international context. Future research could explore how detection systems, investigative tools, and reporting infrastructures integrate into UAV incident response workflows and shape coordination and decision-making. As UAV technologies become more autonomous, further work is also needed to understand how these developments may change investigative methods and incident response processes.

## Acknowledgements

The research in this paper was supported by the U.S. Department of Homeland Security under Award Number 20STTPC00001-06. Any opinions or conclusions contained herein are those of the authors and do not necessarily reflect those of the Department of Homeland Security.